\documentstyle{cjaa}                    
\input{epsf.sty}                        
\input{psfig.sty}                       

\begin{document}

   \title{Intrinsic spectra and energetics of cosmological Gamma--Ray Bursts}

   \author{L. Amati\inst{}\mailto{amati.bo.iasf.cnr.it}} 
   \offprints{L. Amati}                   

  \institute{Istituto di Astrofisica Spaziale e Fisica Cosmica - Sez. Bologna, CNR,
            via P. Gobetti 101, I--40129 Bologna, Italy\\
            \email{amati@bo.iasf.cnr.it}
          }

   \date{Received~~2003 November xx; accepted~~2004~~February xx }

   \abstract{We extend a previous work on the intrinsic spectral properties and 
energetics of GRBs with known redshift based on 12 BeppoSAX events by including in 
the sample a total of 10 more 
events detected either by BATSE, BeppoSAX or HETE--2. An indication of a trend of
the total isotropic equivalent radiated energy, $E_{\rm rad}$, with redshift is found 
and, remarkably, the previously found relationship between the peak energy of the 
rest--frame
$\nu$F$\nu$ spectrum, $E_{\rm p}^{\rm rest}$, and $E_{\rm rad}$ is confirmed and its significance increased.
The implications of these results are briefly discussed in the framework
of standard models for the prompt emission of GRBs. 
   }

   \authorrunning{L. Amati}            
   \titlerunning{Intrinsic spectra and energetics of cosmological Gamma--Ray Bursts}  


   \maketitle
%
%
\section{Introduction}           
\label{sect:intro}

In the past, the spectral studies of GRBs put in evidence very important
properties, like 
fast spectral evolution, non thermal spectrum with a smoothly broken power-law shape
that can be satisfactorily reproduced in many cases by Synchrotron Shock Models (SSM), 
narrow distributions of the spectral parameters, in particular of the peak energy 
$E_{\rm p}$ of 
the $\nu$F$\nu$ spectrum,
hardness-duration and  hardness--intensity relationships (e.g. Band et al. 1993,
Tavani 1997, Paciesas et al. 1999, Preece et al. 2000, Frontera et al. 2000
\nocite{Band93,Tavani97,Paciesas99,Preece00,Frontera00}).
Nevertheless, the impact of these results on the understanding of the nature 
of GRBs and of the emission mechanism(s) producing the prompt emission
were strongly 
limited
by the lack of knowledge of the GRBs distance scale and thus of their energetics and
intrinsic spectral shape.
But in 1997, the BeppoSAX breakthrough 
discovery of afterglow emission from GRBs, leading to the first 
redshift measurements for these sources,
started a new era in GRB science. 
All the GRBs for which a redshift could be estimated lie at cosmological distances 
(z= 0.17 - 4.5) (except for GRB 980425 if associated with SN1998bw, $z$=0.0085)
and the total isotropic equivalent radiated energy, $E_{\rm rad}$, is huge: from $\sim$10$^{51}$ erg 
up to $\sim$10$^{54}$ erg (again with the exception of GRB~980425).
The spectral analysis of GRBs with known redshift ($\sim$30 events up to now) can produce
a big step forward with respect
to previous spectral studies, allowing e.g.
the measurement of the distribution of the rest--frame peak energy, 
$E_{\rm p}^{\rm rest}$, the
study of spectral parameters as a function of 
$E_{\rm rad}$ and redshift, 
the search for redshift indicators.\\
In a previous work (Amati et al. 2002), we studied the intrinsic spectra and 
energetics of a
sample of 12 GRBs simultaneously detected by the WFC (2-28 keV) and GRBM 
(40-700 keV) instruments on board BeppoSAX (Boella et al. 1997 \nocite{Boella97})
and 
for which redshift estimates were available.
This analysis took advantage also of the extension of the spectral fitting energy 
range down to 2 keV, allowing better estimate of low energy index $\alpha$
with respect 
e.g. to BATSE (20-2000 keV) and to reduce the $\alpha$--$E_{\rm p}$ 
correlation and data 
truncation effects. The main results were
the indications of a trend of $E_{\rm rad}$ and $\alpha$ with $z$
and, importantly, the 
evidence of a correlation between $E_{\rm p}^{\rm rest}$ and $E_{\rm rad}$.
The latter correlation is in agreement with previous indications inferred from BATSE 
detected events without knowledge of $z$ (Lloyd, Petrosian \& Mallozzi 2000)
and with the predictions of some 
realizations of SSM scenarios.
In addition, our results indicated that the intrinsic distribution of $E_{\rm p}^{\rm rest}$ could be
 much broader than that inferred from non redshift--corrected spectra.
A detailed analysis of the combined WFC/GRBM sensitivity thresholds showed that 
selection and data truncation effects did not affect significantly our results.\\
The major problem with these results was the small number of events included in the 
sample, but now spectral parameters are available for more GRBs with 
known redshift (e.g. Jimenez, Band \& Piran 2001, Barraud et al. 2003 
\nocite{Jimenez01,Barraud03}). 
In this work we present and discuss the results of the same analysis on
an enlarged sample including, 
in addition to the 12 events considered in Amati et al. (2002), 4 HETE--2 events, 
4 BATSE events and 2 more BeppoSAX events, for a total of 22 events (Table~1).


\section{GRB sample and data analysis}
\label{sect:Obs}

The GRBs included in our sample and their redshift values, 
see Jimenez, Band \& Piran (2001), 
Amati et al. (2002),
Bloom (2003), Atteia (2003) 
\nocite{Jimenez01,Amati02,Bloom03,Atteia03} for a complete list of references, 
are reported in Table~1, together with the most relevant best fitting
parameters of their intrinsic spectra and the computed isotropic equivalent radiated
energies.
The fitting model was either the canonical Band function (Band et al. 1993 
\nocite{Band93}), 
whose spectral parameters are the low energy index $\alpha$, the high energy 
index $\beta$ and the break energy E$_0$,
 or, for
three of the HETE--2 events (see Barraud et al. 2003),
a cut--off power--law with index $\alpha$ and 
cut--off energy E$_c$ . The spectral shape of these two models is nearly the same
up to energies not much higher than $E_{\rm p}^{\rm rest}$, which is given by $(2+\alpha)$ 
multiplied by E$_0$ or E$_c$.
For the BeppoSAX events, the fits were performed on the blue-shifted spectra in order 
to obtain the spectral shape in the GRB cosmological rest-frame, as was done in Amati 
et al. (2002).
For the HETE--2 and BATSE events we derived the parameters of the cosmological 
rest-frame spectra from the published measured ones by accounting for the fact that 
$\alpha$ and $\beta$ are invariant with redshift,  E$_0$ scales as $(1+z)$ and 
the overall 
normalization as $(1+z)^{-\alpha-1}$. 
The measured spectral parameters of
the BATSE events were taken from Jimenez, Band \& Piran (2001) and those of the HETE--2 events 
were taken from
Barraud et al. (2003) and Atteia (2003). The uncertainties on the
parameters values reported in Table~1 are at 1$\sigma$ level and, when not available, 
are conservatively assumed to
be 20\% of the best fit values. 
The total radiated energies were then computed by integrating the best fit models to 
the
cosmological rest--frame spectra from 
1 to 10000 keV and scaling by the luminosity distances in the way described by 
Amati et al. (2002). We assumed a standard cosmology with H$_0$=65 km/s/Mpc, 
$\Omega$$_m$=0.3 and $\Omega$$_\Lambda$=0.7 .
The uncertainties on the $E_{\rm rad}$ values reported in Table~1 were
derived from those 
on the measured fluences, 
which were assumed to be 10\% when not available. The Pearson's correlation 
coefficients $r$
reported and discussed in the next section were performed by properly weighting
for data uncertainties.

\section{Results}
\label{sect:Res}

The spectral parameters in the cosmological rest frame and the
isotropic equivalent total radiated energies of the GRBs included in our sample,
computed as described in the previous section,
are reported in Table~1. The values of E$_{rad}$ span over nearly
three orders of magnitude and a considerable spread of the intrinsic $E_{\rm p}^{\rm rest}$ 
values is found. The values of $\alpha$ also vary substantially from burst
to burst and are always consistent with the limits predicted for optically
thin synchrotron emission ($-$0.67,$-$1.5).
As can be inferred directly from Table~1, the 
indication of a dependence of $\alpha$ on $z$ is slightly weakened with respect to what 
found for
the 'old' sample. In Fig.~1 and Fig.~2 we plot $E_{\rm rad}$ vs. $z$ and 
$E_{\rm p}^{\rm rest}$ vs. $E_{\rm rad}$, respectively.
By adding the new 10 events (triangles), the trend of 
$E_{\rm rad}$ with $z$ is confirmed; 
in particular there is indication of a trend of the maximum isotropic equivalent
radiated energy with redshift, with the only exception of GRB011121. 
The correlation 
coefficient between the logarithms of these two quantities is $r$=0.61 for 22
events, corresponding to a chance probability of $\sim$0.2\% .
But, as apparent from Fig.~2, the most striking result is that with the addition of
the new events 
the correlation between $E_{\rm p}^{\rm rest}$ and $E_{\rm rad}$ is confirmed and its significance 
increased. Indeed, the correlation coefficient between log($E_{\rm p}^{\rm rest}$)
 and log$(E_{\rm rad}$)
is $r$ = 0.90, corresponding, for 20 events (the upper limits available for
GRB000214 and GRB010222 were obviously not included in the correlation analysis)
to a chance probability as low as 1.3$\times$10$^{-7}$ . The slope of this
relationship is 0.35$\pm$0.06, which is lower than
the value found for the 'old' sample (Amati et al. 2002).

\begin{table}[]
  \caption[]{Redshift, intrinsic spectral parameters and energetics of the GRBs
included in our sample. See text for details and references.}
  \label{Tab:tab1}
  \begin{center}\begin{tabular}{ccccccc}
  \hline\noalign{\smallskip}
GRB & Mission  & $z$ & Model & $\alpha$ & $E_{\rm p}^{\rm rest}$ & $E_{\rm rad}$               \\
 &  &  & &  & (keV) & (10$^{52}$ erg)             \\
  \hline\noalign{\smallskip}
970228  & SAX  & 0.695 & Band & $-$1.54$\pm$0.08 &  195$\pm$64  & 1.86$\pm$0.14     \\
970508  & SAX  & 0.835 & Band & $-$1.71$\pm$0.10 &  145$\pm$43  & 0.71$\pm$0.15     \\
970828  & BATSE  & 0.958 & Band & $-$0.70$\pm$0.14 & 586$\pm$117  & 34$\pm$4     \\
971214  & SAX  & 3.412 & Band & $-$0.76$\pm$0.17 &  685$\pm$133  & 24$\pm$3     \\
980326  & SAX  & 0.9--1.1 & Band & $-$1.23$\pm$0.21 & 71$\pm$36  & 0.56$\pm$0.11     \\
980329  & SAX  & 2.0--3.9 & Band & $-$0.64$\pm$0.14 & 935$\pm$150  & 211$\pm$20    \\
980613  & SAX  & 1.096 & Band & $-$1.43$\pm$0.24 &  194$\pm$89  & 0.68$\pm$0.11     \\
980703  & BATSE  & 0.966 & Band &$-$1.31$\pm$0.26  &  503$\pm$64  & 8.3$\pm$0.8     \\
990123  & SAX  & 1.600 & Band & $-$0.89$\pm$0.08   &  2030$\pm$161  & 278$\pm$32     \\
990506  & BATSE  & 1.307 & Band & $-$1.37$\pm$0.28  & 624$\pm$130  & 109$\pm$11     \\
990510  & SAX  & 1.619 & Band & $-$1.23$\pm$0.05   &  423$\pm$42  & 20$\pm$3     \\
990705  & SAX  & 0.842 & Band & $-$1.05$\pm$0.21 &  348$\pm$28  & 21$\pm$3     \\
990712  & SAX  & 0.433 & Band & $-$1.88$\pm$0.07 &  93$\pm$15  & 0.78$\pm$0.15     \\
991216  & BATSE  & 1.02 & Band & $-$1.23$\pm$0.25 &  645$\pm$130  & 50$\pm$5     \\
000214  & SAX  & 0.37--0.47 & CPL & $-$1.62$\pm$0.13  & $>$117  & 0.93$\pm$0.03     \\
010222  & SAX  & 1.477 & Band & $-$1.35$\pm$0.19 &  $>$886  & 154$\pm$17     \\
010921  & HETE--2 & 0.451 & CPL & $-$1.49$\pm$0.05 &  152$\pm$37  & 1.10$\pm$0.11     \\
011121  & SAX  & 0.362 & Band & $-$1.42$\pm$0.14 &  295$\pm$35  & 11$\pm$1     \\
011211  & SAX  & 2.140 & Band & $-$0.84$\pm$0.09 &  186$\pm$24  & 6.3$\pm$0.7     \\
020124  & HETE--2  & 3.2 & CPL & $-$1.10$\pm$0.08 &  504$\pm$95  & 31$\pm$0.3     \\
020813  & HETE--2  & 1.254 & CPL & $-$1.05$\pm$0.02 &  477$\pm$22  & 86$\pm$9     \\
021211  & HETE--2  & 1.01 & Band & $-$0.90$\pm$0.18 &  116$\pm$24  & 0.61$\pm$0.12     \\
  \noalign{\smallskip}\hline
  \end{tabular}\end{center}
\end{table}

\begin{figure}[t]
   \begin{center}
   \centerline{\psfig{figure=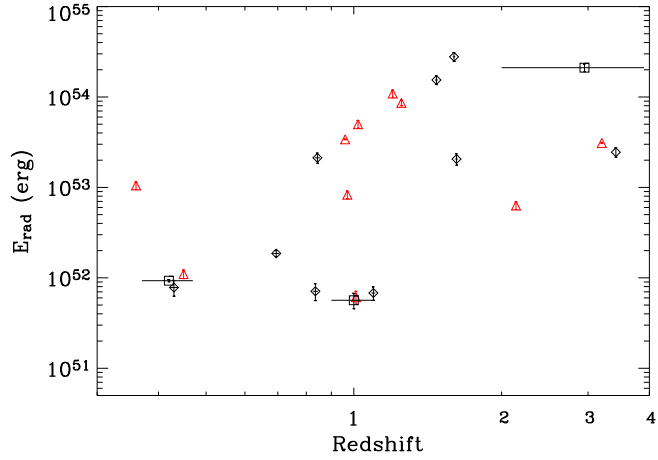,width=9.5cm}}
   \caption{Isotropic equivalent radiated energy as a function of redshift for the
GRBs in our sample. Diamonds (events with firm redshift estimates) and squares (GRB980326, GRB980329 and GRB000214) indicate the events 
included in the sample of Amati et al. (2002), triangles the new events included in our sample.}
   \label{Fig:plot1}
   \end{center}
\end{figure}
\begin{figure}[t]
   \begin{center}
   \centerline{\psfig{figure=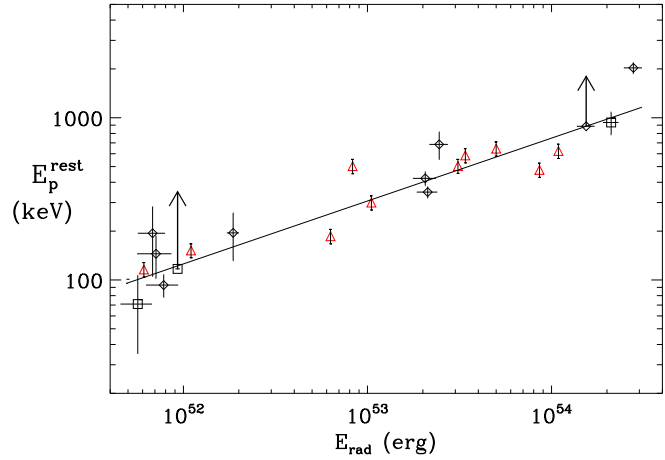,width=9.5cm}}
   \caption{Peak energy of the intrinsic $\nu$F$\nu$ spectrum as a function of 
isotropic equivalent radiated energy for the
GRBs in our sample. Diamonds and squares (GRB980326, GRB980329 and GRB000214) indicate the events included in the sample of Amati et al. (2002), triangles the new events included in our sample.}
   \label{Fig:plot2}
   \end{center}
\end{figure}

\section{Discussion}
\label{sect:analysis}

With respect to our previous work (Amati et al. 2002), the study reported in this
paper is based on a 
much larger sample (22 against 12 events) and thus the significance of the
results is substantially increased. For a discussion of possible selection and
data truncation effects, indicating that they should not affect
significantly our results, see Amati et al. (2002), Jimenez, Band \& Piran (2001) and
Barraud et al. (2003) for the BeppoSAX, BATSE and HETE--2 events, respectively.\\
The trend of $E_{\rm rad}$ with $z$
may indicate an higher energy budget for older events. In this case, the high 
dispersion may reflect significant variation in shock parameters, 
and thus in shock radiating efficiency, and/or in beaming angles, in the case that the
assumption of isotropic emission is not valid for several events in our sample.
If the emission of most GRBs is collimated and the radiated energy is nearly constant,
as supposed e.g. by Frail et al.(2001) \nocite{Frail01}, then the observed trend of the 
{\it maximum}
$E_{\rm rad}$ with $z$ may 
be due to a combination of the facts that the estimated isotropic equivalent
radiated energy is larger for 
events with smaller jet opening angles and that it is more probable to detect 
events with 
small jet opening angles at high $z$ than at low $z$.
The hypothesis that older GRBs are more energetic is still valid if we assume that
the distribution of jet 
angles is very narrow.

The reinforced evidence of a correlation between $E_{\rm p}^{\rm rest}$ and 
$E_{\rm rad}$
confirms the indications emerged from the observed hardness--intensity correlation 
and previous correlation studies on burst energy break, peak flux and fluence 
performed on events with unknown $z$. For example, 
Lloyd, Petrosian \& Mallozzi (2000) \nocite{Lloyd00}, 
basing on a BATSE sample, inferred log$(E_{\rm p}^{\rm rest})\propto d \times$ log$(E_{\rm rad})$, with 
$0.4<d<0.7$, which is consistent with 
our findings.
The $E_{\rm p}^{\rm rest}$ -- $E_{\rm rad}$ relationship can put constraints on GRB prompt 
emission models, like SSM internal shocks, Inverse Compton (IC) dominated 
internal shocks, external shocks, innermost models (see Zhang \& Meszaros 2002 
\nocite{Zhang02} for
a review). In general, $E_{\rm p}^{\rm rest}$ and $E_{\rm rad}$ are both quantities
dependent on the fireball bulk Lorentz factor $\Gamma$, in a way
varying for each scenario and for different assumptions 
(e.g. Schaefer 2003 \nocite{Schaefer03}). For instance, as pointed out by
Zhang \& Meszaros (2002): a) the correlation we found 
implies for SSM internal shocks that the slope of the power--law linking
$E_{\rm p}^{\rm rest}$ and $\Gamma$ is 
$<1/4$, unless invoking relevant IC and/or peculiar assumptions; b)
external shocks 
predict a steeper slope of the $E_{\rm p}^{\rm rest}$ -- $E_{\rm rad}$ power--law relationship
with respect to what we found; c)
innermost (i.e. baryon 
photosphere) emission models predict a slope consistent with our results, but this 
mechanism is expected to contribute only a fraction of the total GRB prompt emission. 

If the emission of the GRBs included in our sample is collimated, then the spread of 
jet angles, if not too large, may contribute to 
the scatter around the best fit power-law of the $E_{\rm p}^{\rm rest}$ -- $E_{\rm rad}$
relationship. By assuming, as above, that most events are 
jetted and radiate a nearly constant energy, then more collimated events have an higher 
equivalent isotropic radiated energy and thus the true correlation may be between 
$E_{\rm p}^{\rm rest}$ and the jet 
opening angle $\theta$$_{\rm j}$ (and the relation between $\theta$$_{\rm j}$ 
and $\Gamma$ 
would also play a role).

The constraints coming from the $E_{\rm p}^{\rm rest}$ -- $E_{\rm rad}$ relationship
have to be combined with those 
coming  from the broadness of the $E_{\rm p}^{\rm rest}$ distribution. Indeed, our results 
indicate 
that the 
intrinsic $E_{\rm p}^{\rm rest}$ distribution is broader than inferred from e.g. BATSE bright 
bursts 
(Preece et al. 2000 \nocite{Preece00}), and thus is less critical for the 
models 
(the broadness 
increases with the number of independent parameters of the model, e.g. 
Zhang \& Meszaros 2002). 

Finally, the estimates of $E_{\rm p}^{\rm rest}$ for further  
HETE-2 GRBs with known redshift, remarkably including the X--Ray Flash 020903 (if its
redshift estimate is true), confirm and extend the $E_{\rm p}^{\rm rest}$ -- $E_{\rm rad}$
relationship (Lamb, Donaghy \& Graziani 2004 \nocite{Lamb04}).

\label{lastpage}

\end{document}